\documentclass{article}

\usepackage{arxiv}

\usepackage[numbers, compress]{natbib}
\usepackage[utf8]{inputenc} 
\usepackage[T1]{fontenc}    
\usepackage{hyperref}       
\usepackage{url}            
\usepackage{booktabs}       
\usepackage{amsmath}        
\usepackage{amsfonts}       
\usepackage{nicefrac}       
\usepackage{microtype}      
\usepackage{xcolor}         
\usepackage{subcaption}     
\usepackage{layouts}        

\hypersetup{
    pdfborder={0 0 0},
    colorlinks=true,
    urlcolor=blue,
    linkcolor=blue,
    citecolor=blue,
    bookmarks=true,
    bookmarksnumbered=true
}

\usepackage{tikz}
\usetikzlibrary{decorations.pathreplacing, positioning, shapes, arrows.meta}

\newcommand{\A}{\mathbf{A}}
\newcommand{\B}{\mathbf{B}}
\newcommand{\Q}{\mathbf{Q}}
\newcommand{\R}{\mathbf{R}}
\newcommand{\h}{\mathbf{h}}
\newcommand{\w}{\mathbf{w}}

\renewcommand{\v}{\mathbf{v}}
\newcommand{\etabf}{\boldsymbol{\eta}}

\newcommand{\mubf}{\boldsymbol{\mu}}
\newcommand{\sigmabf}{\boldsymbol{\sigma}}

\def\approxprop{%
    \def\p{%
        \setbox0=\vbox{\hbox{$\propto$}}%
        \ht0=0.6ex \box0 }%
    \def\s{%
        \vbox{\hbox{$\sim$}}%
    }%
    \mathrel{\raisebox{0.7ex}{%
            \mbox{$\underset{\s}{\p}$}%
    }}%
}

\title{Towards Data Assimilation in Level-Set Wildfire Models Using Bayesian Filtering}

\date{}

\author{%
    Joel Janek Dabrowski$^{\dagger}$, 
    Carolyn Huston, 
    James Hilton, 
    St\'{e}phane Mangeon, 
    Petra Kuhnert \\
    Data61, CSIRO \\
    Australia \\
    \texttt{Firstname.Lastname@data61.csiro.au}
}

\begin{document}

\maketitle

\begin{abstract}
The level-set method is a prominent approach to modelling the evolution of a fire over time based on a characterised rate of spread. It however does not provide a direct means for assimilating new data and quantifying uncertainty. Fire front predictions can be more accurate and agile if the models are able to assimilate data in real time. Furthermore, uncertainty estimation of the location and spread of the fire is critical for decision making. Using Bayesian filtering approaches, we extend the level-set method to allow for data assimilation and uncertainty quantification. We demonstrate these approaches on data from a controlled fire.
\end{abstract}



\section{Introduction}

Wildfires are highly destructive and have significant costs associated with them \cite{miller2015spark}. These costs may be associated with damaged infrastructure, destroyed crops, loss of wildlife, and even loss of human life. A wildfire model is a critical tool for emergency planners and decision makers. It provides the means to understand how a wildfire is developing and how it may progress. Wildfire models are characterised by inputs such as fuel type, wind, and topography, which can be highly variable and uncertain. For example, wind has a high temporal and spatial variance, thereby having the potential to generate a broad range of fire front shapes with variable directions. Data assimilation provides the means to update the model in real time given observations of the actual fire front. 

The level-set method forms the basis of prominent wildfire simulation platforms such as Spark \cite{miller2015spark}\footnote{For example, see \url{https://www.csiro.au/en/research/technology-space/ai/spark}}, but it does not provide any direct means for data assimilation and uncertainty estimation. Data assimilation and uncertainty estimation have been considered in wildfire models \cite{Srivas2016Wildfire,Xue2012Data,Rochoux2015Towards, Silva2014Application}. However, to our knowledge, other than \citet{Rochoux2013Regional} -- who use the Best Linear Unbiased Estimator (BLUE) to assimilate instantaneous snapshots of the fire front positions into a level-set model -- assimilation and uncertainty estimation has not been incorporated into the level-set wildfire models. Our contribution thus includes a preliminary investigation into how Bayesian filtering approaches can be used to incorporate data assimilation and uncertainty estimation into level-set wildfire models.

\section{Methods}

\subsection{Rate of Spread Models}
\label{sec:fireModelling}

The Rothermel model \cite{rothermel1972mathematical} is a popular rate of spread (ROS) model for fires. Excluding topography, Hilton et al. \cite{hilton2016curvature,hilton2015effects} represented it as
\begin{align}
    \label{eq:fireSpreadModel}
    s = \beta + \gamma \omega
\end{align}
where $s$ is the ROS, $\beta$ is a constant describing the characteristic fire front speed for a straight line fire burning in the absence of wind, $\gamma$ is a growth parameter due to a constant wind speed, and $\omega$ is the wind speed in the direction normal to the fire front. In this study, the parameters $\beta$ and $\gamma$ are treated as constants (assuming constant and homogenous combustion conditions) and are manually calculated from the data. For example, $\beta$ can be estimated from the average distance the fire spreads in the direction normal to the wind and $\gamma$ can be estimated from the average distance the fire spreads in the direction of the wind.

\subsection{Level-Set Method}

The ROS model describes how a fire spreads in a single dimension. To model how the fire spreads in a two-dimensional space, the level-set method \cite{Osher1998Fronts} can be used \cite{mallet2009modeling,hilton2016curvature}. The level-set method represents a fire front as a closed contour, using an auxiliary three-dimensional surface called the level-set function. The closed contour fire front is defined as the intersection between the level-set function and the zero-plane, and is called the zero level-set. Using a three-dimensional surface is beneficial as it can handle changes in topology such as shape splits and mergers \cite{Osher2001Level}. 

The level-set equation in its first form is a Hamilton–Jacobi equation and is given by \cite{hilton2016curvature,strang2007computational,Osher2001Level}
\begin{align}
    \label{eq:levelSetEquation}
    \frac{\partial \phi}{\partial t} + u | \nabla \phi | = 0
\end{align}
where $\phi(x,y,t)$ is the level-set function in the form of the signed distance function, which describes the distance to the zero level-set at time $t$ and location $(x,y)$. The speed $u$ is the normal component of the velocity at the interface. In the wildfire application, the zero level-set $\phi(x,y,t) = 0$ represents the fire front, which moves normal to itself at speed $u$. The speed $u$ can be directly described using a ROS model such as (\ref{eq:fireSpreadModel}). The level-set equation is solved using an accurate finite difference numerical method such as the Essentially Nonoscillatory (ENO) method \cite{osher2003level}.


\subsection{Bayesian Filtering}

Bayesian filtering is an approach to infer internal latent states of some dynamical system based on partial noisy observations. Given some latent variable $\h_t$ at time $t$ and an observed variable $\v_t$, the dynamical system is described by \cite{murphy2012machine}
\begin{align}
    \label{eq:stateTransition}
    & \h_t = f(\h_{t-1}) + \etabf^{h} \\
    \label{eq:measurement}
    & \v_t = g(\h_t) + \etabf^{v}
\end{align}
where $f$ is the state transition function, $g$ is the measurement function, and $\etabf^{h} \sim \mathcal{N}(\mathbf{0}, \Q)$ and $\etabf^{v} \sim \mathcal{N}(\mathbf{0}, \R)$ are noise components. The Bayesian filter estimates the posterior filtered distribution $p(\h_t|\v_{1:t})$, which is the probability distribution of the current latent state given all previous observations. Other than the linear-Gaussian case, $p(\h_t|\v_{1:t})$ is generally intractable. Approximate inference methods such as the particle filter \cite{doucet2013sequential} and the Ensemble Kalman Filter (EnKF) \cite{evensen2006data} are required.

The Ensemble Kalman Filter propagates a set of Monte Carlo samples according to nonlinear dynamics and infers Gaussian parameters from the propagated samples. The particle filter is a sequential Monte Carlo approach that does not assume any parametric form by directly representing the filtered posterior distribution with a set of particles. The particles are propagated over time based on (\ref{eq:stateTransition}) and are updated based on measurements given by (\ref{eq:measurement}). The updates are performed using sequential importance resampling, which provide a means to perform data assimilation in the dynamical model. 

\section{Demonstration}

\subsection{Dataset}
\label{sec:dataset}

\begin{figure}[!t]
    \centering
    \begin{subfigure}[h]{2.0in}
        \includegraphics[]{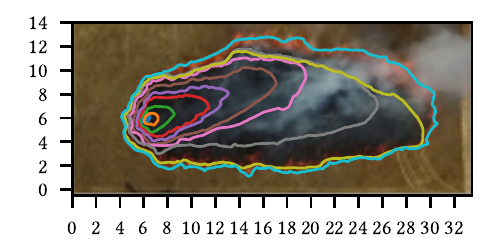}
        \caption{}
        \label{fig:dataset}
    \end{subfigure}
    \begin{subfigure}[h]{2.0in}
        \includegraphics[]{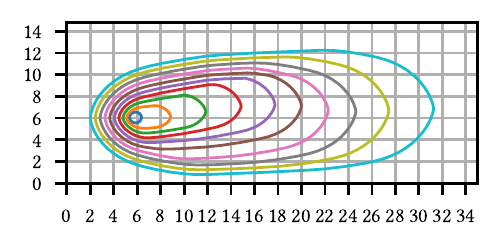}
        \caption{}
        \label{fig:levelSetResults}
    \end{subfigure}
    \begin{subfigure}[h]{2.0in}
        \includegraphics[]{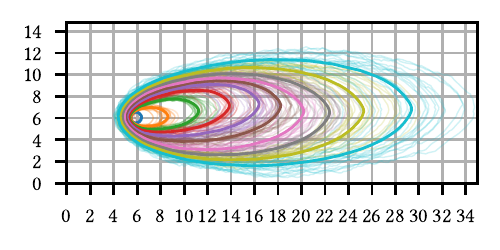}
        \caption{}
        \label{fig:braidwood_pf_levelset}
    \end{subfigure}
    \begin{subfigure}[h]{2.0in}
        \includegraphics[]{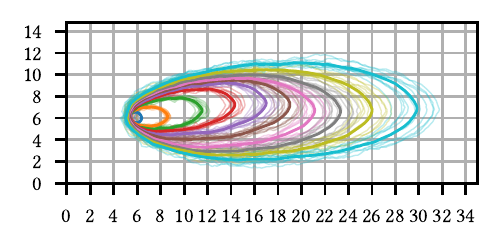}
        \caption{}
        \label{fig:braidwood_enkf}
    \end{subfigure}
    \caption{Plots of the dataset and simulations for 90 seconds at 10 second increments. Each contour plots the fire front at a time increment. The dataset with the manually labelled fire front and a photograph of the fire at 90 seconds is plotted in (a). The level-set simulation is plotted in (b). The particle filter level-set simulation is plotted in (c). The EnKF simulation is plotted in (d). In \figurename{}s (c) and (d), the thick contours are the average of the particle contours at the particular time and the fine lighter contours are the Monte Carlo samples for a particular time. These indicate uncertainty.}
    \label{fig:results}
\end{figure}

The Braidwood fire dataset \cite{sullivan2018study} is used in this study. The fire was a controlled burn that was filmed overhead for over a period of 90 seconds in Braidwood, New South Wales, Australia. The fire front was manually mapped at 10 second intervals. This mapping can be automated in practice as fires are highly observable. Fire observations include hotspots -- geolocated points identified as a fire, usually via satellites (e.g. VIIRS \cite{Schroeder2014New}) -- and more detailed observations such as the fire line or full fire perimeter identified from thermal imaging \cite{jones2009innovations}. Also included in the dataset are the wind speed and direction for every second that the fire burned. An image of the fire at 90 seconds and contour plots of the fire fronts at every 10 seconds are illustrated in \figurename{~\ref{fig:dataset}}. The fire perimeters are used as the ground truth and the wind velocity is used in the fire spread model defined in (\ref{eq:fireSpreadModel}).

\subsection{Level-Set Method}
For the purpose of a base-line, the Braidwood fire was simulated using the level-set method with an initial circular fire perimeter and using the ROS model given by (\ref{eq:fireSpreadModel}). The simulated fire front is illustrated in \figurename{~\ref{fig:levelSetResults}}. With constant and homogeneous combustion conditions assumed, the shape of the perimeter is very smooth and consistent. The fire front grows outwards at a rate determined by $\beta$, causing the circular perimeter to increase in radius. The fire then spreads in the direction of the wind according to wind parameter $\gamma$, causing the circular fire perimeter to expand in that direction.

\subsection{Combined Particle Filter and Level-Set Method}
\label{sec:combinedMethod}

We apply the particle filtering approach to the Level-Set method such that each particle represents a Level-Set function. A particle's state transition function $f$ in (\ref{eq:stateTransition}) is represented using the level-set equation given by (\ref{eq:levelSetEquation}). Each particle is then propagated by a unique level-set function. The particles are initialised around the location where the fire starts. To achieve variation in the particles, the parameters $\beta$ and $\gamma$ are also initially sampled from Gaussian distributions $\mathcal{N}(\mubf_\beta, \sigmabf_\beta)$ and $\mathcal{N}(\mubf_\gamma, \sigmabf_\gamma)$ respectively. The means $\mubf_\beta$ and $\mubf_\gamma$ are the $\beta$ and $\gamma$ values used in the standard Level-Set method. The standard deviations $\sigmabf_\beta$ and $\sigmabf_\gamma$ represent the initial uncertainty in the parameter values. 

As is often customary in Bayesian filtering applications, the parameters are manually tuned to the particular problem. Initial parameters are important to get the initial variation in the particles, but their affects generally do not propagate far into the future.

The particle filter runs by performing several update steps based on the Level-Set method until a measurement observation arrives. The measurements are the ground-truth fire fronts provided every 10 seconds. Sequential Importance Resampling is used to resample the particles at each update step. This resampling is performed according to the weights $\w \propto p(\v_t | \h_t)$, which provides an indication on how well the observation fits the model's prediction. In general, measuring the accuracy of a fire-front contour is however challenging \cite{filippi2013representation}. In our approach we take advantage of the discrete spatial grid that the Level-Set method is evaluated over. The fire-front contours can be treated as greyscale images in two-dimensional space. The measurement noise is approximated by applying a Gaussian filter to the contours which blurs the contours in image space. This filtered image provides an approximation of the measurement in (\ref{eq:measurement}) and the ground-truth fire fronts can be evaluated according to corresponding pixel values. That is
\begin{align*}
    p(\v_t | \h_t) \approxprop \sum_i^M \sum_j^N A_{i,j} B_{i,j}
\end{align*}
The matrix $\A$ is the image holding the ground-truth fire front contour, where the contour is represented with a value of $1$ and the remaining pixels are $0$. The matrix $\B$ is the image with the predicted fire front contour, which has been burred with a Gaussian filter.

The simulation results are illustrated in \figurename{~\ref{fig:braidwood_pf_levelset}}. For this simulation, 100 particles were used. Each particle level-set expands and is affected by the wind differently according to its parameters. The white process noise $\etabf^h \sim \mathcal{N}(\mathbf{0}, \Q_t)$ tends to perturb the fire front perimeter. The contours in the predictions are the average of the particles at that time. An uncertainty representation can be provided by calculating the variance over the particles. In \figurename{~\ref{fig:braidwood_pf_levelset}} is visualised by plotting 20 of the particles. These results show an improvement compared with the level-set results in \figurename{~\ref{fig:levelSetResults}}.

\subsection{Dynamic Updating of Parameters via the Ensemble Kalman Filter}
The combined particle filter and level-set method can be computationally expensive as each particle is a Level-Set equation. An alternative approach is to use the Ensemble Kalman filter (EnKF) to dynamically update the parameters $\beta$ and $\gamma$ in a single level-set ROS model. The parameters $\beta$ and $\gamma$ are treated as the latent variables and $f$ in (\ref{eq:stateTransition}) is defined such that 
%
\begin{align*}
    \h_t = \h_{t-1} + \etabf^{h}
\end{align*}
That is, the parameters are assumed to change slowly over time. The measurement function $g$ in (\ref{eq:measurement}) is the level-set equation. The observation $\v_t$ is thus the fire front obtained by propagating the level-set equation given the current parameters in $\h_t$. To compute the residual in the EnKF, fire front contours are represented as images as discussed in Section (\ref{sec:combinedMethod}).

The simulation for the EnKF is illustrated in \figurename{\ref{fig:braidwood_enkf}}. A total of 10 EnKF samples were used. Each contour in the prediction is the Monte Carlo mean. An uncertainty representation can be provided by calculating the variance over the samples. In \figurename{~\ref{fig:braidwood_enkf}} is visualised by plotting the 10 EnKF samples. The results are less accurate than the other methods as the fire fronts have expanded too quickly and tuning the EnKF proved to be challenging. It is expected that this approach could be improved with an alternative residual measure in the EnKF. 


\section{Discussion and Conclusion}
\label{sec:conclusion}

The combined particle filter and level-set method provides the most promising results. This method is however computationally expensive as it requires propagating several level-set equations as particles. The propagation of particles could however be computed in parallel. The EnKF does have potential, but more work is required to make the EnKF more stable.

A striking feature is that the contours in both \figurename{~\ref{fig:braidwood_pf_levelset}} and \figurename{~\ref{fig:braidwood_enkf}} become more elliptical in shape as time progresses. Fire fronts are well known to be elliptical in nature (e.g. \cite{anderson1982modelling}), and both of the Bayesian filtering approaches seem to capture this. The key addition was including noise in the Rothermel model equations.

The level-set method is required to use a small time step in the finite difference approximation. The result is that many prediction steps are required between the 10 second data sample intervals. The predictions can begin to diverge from actual fire front due to inaccuracies in the models and data noise. The result is that the update steps (data assimilation) may not be as effective. Having a higher frequency of data samples would improve the results and stability of the models.

A common problem encountered by all approaches is that the fire seems to accelerate in the direction of the wind, as is indicated by the uneven spacing between the contours. The models however assume a constant ROS. It is possible that this fire did not reach steady state. In future work, we intend to improve on the EnKF approach, develop more automatic approaches to fitting the models to the data (and testing on other datasets), and investigate using more complex ROS models.

\section*{Acknowledgements}
This work is supported by the Machine Learning and Artificial Intelligence Future Science Platform at CSIRO.

\begin{small}
    \bibliographystyle{unsrtnat}
    \bibliography{Bibliography}
\end{small}

\end{document}